# Improving the methods of email classification based on words ontology

Foruzan Kiamarzpour[1], Rouhollah Dianat[2], Mohammad bahrani[3], Mehdi Sadeghzadeh[4]

[1] Department of Computer Engineering, Science and Research Branch of Bushehr,
Islamic Azad University, Bushehr, Iran
*Foruzan.kiamarzpour@yahoo.com*

[2] Department of Computer Engineering, Faculty of Engineering University,
Qom University, Qom, Iran
*rouhollah.dianat@gmail.com*

[3] Department of Computer Engineering, Sharif University of Technology, Tehran, Iran
*bahrani@ce.sharif.edu*

[4] Department of Computer Engineering, Faculty Member,
Azad Islamic University of Mahshahr,Mahshahr, Iran
*sadegh_1999@yahoo.com*

**Abstract**

The Internet has dramatically changed the relationship among people and their relationships with others people and made the valuable information available for the users. Email is the service, which the Internet provides today for its own users; this service has attracted most of the users' attention due to the low cost. Along with the numerous benefits of Email, one of the weaknesses of this service is that the number of received emails is continually being enhanced, thus the ways are needed to automatically filter these disturbing letters. Most of these filters utilize a combination of several techniques such as the Black or white List, using the keywords and so on in order to identify the spam more accurately In this paper, we introduce a new method to classify the spam. We are seeking to increase the accuracy of Email classification by combining the output of several decision trees and the concept of ontology.

**Keywords**: *Data Mining, Email classification, spam detection, decision tree, ontology* .

## 1 . Introduction

Unwanted spam is sent to the users with different purposes. According to the estimates by Ferris in his research [1], 15 to 20% of emails are spam letters and about half of the users are receiving 10 or more spam every day. Sometimes, some of them receive more than a few hundred unwanted emails. International Data Group [2] has estimated that the traffic of email is more than 60 billion messages daily by 2006.

Nowadays, most of the measures, done on the spam filtering, are based on the techniques such as classifications of Naïve

Bayesian, neural networks, and.... We have provided a framework for efficient spam filtering by using the ontology. Ontology makes sense for the machines to understand the meaning of data [3]. This paper represents an efficient spam filtering method by using the ontology. We have used the SpamBase Dataset[4] and Weka and Jena in order to make the ontology. Using ontology that is specially designed to filter spam, bunch of unsolicited bulk email could be filtered out on the system.

This paper proposes to find an efficient spam email filtering method using ontology. We used Waikato Environment for Knowledge Analysis (Weka) explorer, and Jena to make ontology based on sample dataset.

Ontology is a kind of computational model of the parts in the world and is often shown by the help of semantic networks.

The ontology is in fact an agreement on a common conceptualization and contains the frameworks for modeling the domain knowledge and the agreements in the field of how some of its theories are represented. [5]

Using the meaning of data, the use of electronic technology becomes much easier. There are multiple languages such as XML, RDF, RDF-schema (RDFS), DAML+OIL, and OWL in order to make the ontology and numerous tools for development and implementation of ontology metadata by using these languages .

We will have the following sections in this paper: In Section 2, the research background and related works are presented; in Section 3, the text classification is described; in section 4, the article idea is provided by using the ontology; and in Section 5 the experiments have been presented by using the proposed framework and Finally, in Section 6 we have mentioned the conclusion and future works.

## 2 . Research background and related works

Some of the conducted studies in the field of Email filtering are described as follows:

Recently the evolution of the Web has attracted interest in defining features, signals for ranking [6] and spam filtering [7, 8, 9, 10, 11]. The earliest results investigate the changes of Web content with the primary interest of keeping a search engine index up-to-date [16, 17].

Many researchers also suggested new systems using other methods such as Bayesian network enhancing the performance of Bayesian classification[12], and WBC (Weighted Bayesian Classifier) that gives weight on some key terms that representing the content's class by SVM(Support Vector Machine)[13]

In this paper we used an ontology for Email Classification ,In additional, j48, ADTree and LADTree was used to classify training dataset.

## 3. Text Classification

Text classification uses the data mining techniques. Most of the data mining tasks are based on the databases and data structures. Some of these methods are described as follows.

### 3.1 Support Vector Machines(SVM)

Support Vector Machines are supervised learning methods used for classification, as well as regression. The advantage of Support Vector Machines is that they can make use of certain kernels in order to transform the problem, such that we can apply linear classification techniques to non-linear data. Applying the kernel equations arranges the data instances in such a way within the multi-dimensional space, that there is a hyper-plane that separates data instances of one kind from those of another.

The kernel equations may be any function that transforms the linearly non-separable data in one domain into another domain where the instances become linearly separable. Kernel equations may be linear, quadratic, Gaussian, or anything else that achieves this particular purpose.

### 3.2 Naïve Bayes Classifier

The Naïve Bayes classifier works on a simple, but comparatively intuitive concept. Also, in some cases it is also seen that Naïve Bayes outperforms many other comparatively complex algorithms. It makes use of the variables contained in the data sample, by observing them individually, independent of each other.

The Naïve Bayes classifier is based on the Bayes rule of conditional probability. It makes use of all the attributes contained in the data, and analyses them individually as though they are equally important and independent of each other.

### 3.3 J48 Decision Trees

A decision tree is a predictive machine-learning model that decides the target value (dependent variable) of a new sample based on various attribute values of the available data. The internal nodes of a decision tree denote the different attributes, the branches between the nodes tell us the possible values that these attributes can have in the observed samples, while the terminal nodes tell us the final value (classification) of the dependent variable.

The attribute that is to be predicted is known as the dependent variable, since its value depends upon, or is decided by, the values of all the other attributes. The other attributes, which help in predicting the value of the dependent variable, are known as the independent variables in the dataset.

### 3.4 Alternating Decision Trees(ADTree)

ADTree [14] is a boosted DT. An ADTree consists of prediction nodes and splitter nodes. The splitter nodes are defined by an algorithm test, as, for instance, in C4.5, whereas a prediction node is defined by a single value x ? R2. In a standard tree like C4.5, a set of attributes will follow a path from the root to a leaf according to the attribute values of the set, with the leaf representing the classification of the set. In an ADTree, the process is similar but there are no leaves. The classification is obtained by the sign of the sum of all prediction nodes existing in the path. Different from standard trees, a path in an ADTree begins at a prediction node and ends in a prediction node.

### 3.5 LADTree

LADTree [15] produces an ADTree capable of dealing with data sets containing more than two classes. The original formulation of the ADTree restricted it to binary classification problems; the LADTree algorithm extends the ADTree algorithm to the multi-class case by splitting the problem into several two-class problems.

Logical Analysis of Data is the method for classification proposed in optimization literature. It builds a classifier for binary target variable based on learning a logical expression that can distinguish between positive and negative samples in a data set. The basic assumption of LAD model is that a binary point covered by some positive patterns, but not covered by any negative pattern is positive, and similarly, a binary point covered by some negative patterns, but not covered by positive pattern is negative. The construction of Lad model for a given data set typically involves the generation of large set patterns and the selection of a subset of them that satisfies the above assumption such that each pattern in the model satisfies certain requirements in terms of prevalence and homogeneity.

# 4 . Spam filtering by using the ontology

## 4.1 General Approach

The first step is to make a smart decision tree, and then we obtain the ontology based on the classification of trees j48, AD and LAD and use the format RDF based on the object, subject, and prediction for creating the ontology.

The second step is to map the decision tree to the ontology and then get a query from the obtained ontology and give it a test Email and determine whether it is spam or not.

First, we should collect a good database. These data should consider the features of a valid email and junk mail. In this paper, we have used the database spambase [4] which contains 4601 emails of which 39.4% are the spam and 60.6% are the valid emails. We have evaluated a number of decision trees in Weka and have decided to use the trees j48, AD and LAD. The input format of data has the format Arff; we convert the training data into the format Arff and build a decision tree based on the training data. All leaves have the values equal to 0 or 1 in the classification and if it is 0, the email is valid and if is 1, it is the spam. This decision tree is a kind of ontology.

## 4.2 Architecture and Implementation

The following figure shows our proposed framework for filtering the spam. The training package is a set of emails which provide the result of classification for us. We have classified the training package by using the decision trees j48, AD, and LAD. As a result of data test, a new email entered system, is properly classified and it is determined whether it is spam or not.

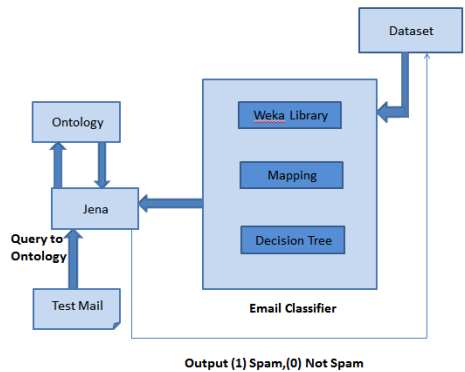

Fig. 1 The architecture of spam filtering by the help of ontology

Therefore, we create the decision trees by using the software Weka and since the decision tree is considered as a kind of ontology, we convert it into the proper format of ontology and ensure its accuracy; we have chosen RDF for creating the ontology (Figure 5)

```
word_freq_our <= 0.71
|   word_freq_your <= 1.04: 0 (824.0/29.0)
|   word_freq_your > 1.04
|   |   word_freq_technology <= 0.04
|   |   |   capital_run_length_longest <= 10: 0 (87.0/5.0)
|   |   |   capital_run_length_longest > 10
|   |   |   |   word_freq_over <= 1.12
|   |   |   |   |   char_freq_( <= 0.064
|   |   |   |   |   |   char_freq_! <= 0.154
|   |   |   |   |   |   |   word_freq_will <= 0.95: 0 (8.0)
|   |   |   |   |   |   |   word_freq_will > 0.95
|   |   |   |   |   |   |   |   word_freq_will <= 1.71: 1 (4.0)
|   |   |   |   |   |   |   |   word_freq_will > 1.71: 0 (4.0/1.0)
|   |   |   |   |   |   char_freq_! > 0.154: 1 (2.0)
|   |   |   |   |   char_freq_( > 0.064: 0 (16.0)
|   |   |   |   word_freq_over > 1.12: 1 (2.0)
|   |   word_freq_technology > 0.04
|   |   |   word_freq_technology <= 0.54: 0 (4.0)
|   |   |   word_freq_technology > 0.54: 1 (3.0)
word_freq_our > 0.71
```

Fig. 2 Part of j48 Classification Result

Figure 2 shows how we choose the J48 classification filter, which uses the simple c4.5 decision tree for calassification.

```
<rdf:RDF
    xmlns:rdf="http://www.w3.org/1999/02/22-rdf-syntax-ns#"
    xmlns:eeg="http://relation/" >
  <rdf:Description rdf:about="http://DecisionTree/Test34">
    <eeg:isElementOf rdf:resource="http://DecisionTree/Test34/Output"/>
    <eeg:value>0.43</eeg:value>
    <eeg:operator>=</eeg:operator>
    <eeg:attribute>word_freq_our</eeg:attribute>
  </rdf:Description>
  <rdf:Description rdf:about="http://DecisionTree/Test85">
    <eeg:isElementOf rdf:resource="http://DecisionTree/Test85/Output"/>
    <eeg:value>1.17</eeg:value>
    <eeg:operator>=</eeg:operator>
    <eeg:attribute>word_freq_our</eeg:attribute>
  </rdf:Description>
  <rdf:Description rdf:about="http://DecisionTree/Test60/Output">
    <eeg:info>(0.0)</eeg:info>
    <eeg:decision>28</eeg:decision>
  </rdf:Description>
  <rdf:Description rdf:about="http://DecisionTree/Test6">
    <eeg:isElementOf rdf:resource="http://DecisionTree/Test6/Output"/>
    <eeg:value>0.11</eeg:value>
    <eeg:operator>=</eeg:operator>
    <eeg:attribute>word_freq_our</eeg:attribute>
  </rdf:Description>
  <rdf:Description rdf:about="http://DecisionTree/Test16">
    <eeg:isElementOf rdf:resource="http://DecisionTree/Test16/Output"/>
    <eeg:value>0.24</eeg:value>
    <eeg:operator>=</eeg:operator>
```

Fig. 3  RDF File made for the result of classifying the tree J48

Figure 3 shows the RDF file created based on J48 classification result. The RDF file was used as an input to Jena to create an ontology which will be used to check if the test email is spam or not.

Your RDF document validated successfully.

**Triples of the Data Model**

| Number | Subject | Predicate | Object |
|---|---|---|---|
| 1 | http://DecisionTree/Test34 | http://relation/isElementOf | http://DecisionTree/Test34/Output |
| 2 | http://DecisionTree/Test34 | http://relation/value | "0.43" |
| 3 | http://DecisionTree/Test34 | http://relation/operator | "=" |
| 4 | http://DecisionTree/Test34 | http://relation/attribute | "word_freq_our" |
| 5 | http://DecisionTree/Test85 | http://relation/isElementOf | http://DecisionTree/Test85/Output |
| 6 | http://DecisionTree/Test85 | http://relation/value | "1.17" |
| 7 | http://DecisionTree/Test85 | http://relation/operator | "=" |
| 8 | http://DecisionTree/Test85 | http://relation/attribute | "word_freq_our" |
| 9 | http://DecisionTree/Test60/Output | http://relation/info | "(0.0)" |
| 10 | http://DecisionTree/Test60/Output | http://relation/decision | "28" |
| 11 | http://DecisionTree/Test6 | http://relation/isElementOf | http://DecisionTree/Test6/Output |
| 12 | http://DecisionTree/Test6 | http://relation/value | "0.11" |
| 13 | http://DecisionTree/Test6 | http://relation/operator | "=" |
| 14 | http://DecisionTree/Test6 | http://relation/attribute | "word_freq_our" |
| 15 | http://DecisionTree/Test16 | http://relation/isElementOf | http://DecisionTree/Test16/Output |
| 16 | http://DecisionTree/Test16 | http://relation/value | "0.24" |

Fig. 4 Ternaries of the model RDF

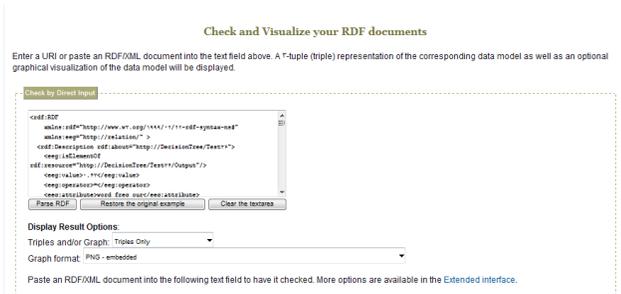

Fig.5 W3C RDF Validation Services

Figure 5 shows RDF validation services. W3C RDF validation services help us to check whether the RDF schema which we are going to give as input to Jena is syntactically correct or not.

## 5 . experiments

We have used 4101 emails in training session and have built the trees by the help of software Weka and converted them to the ontology format. We get the query from this ontology for the test stage and give them 500 test emails for classifying into two groups of spam and valid emails.

The results of processing and all features are shown in the following table; first we investigate the spam class (1) and the valid email class (0):

Definitions of cases mentioned in the next tables and charts:

Tp: The spam, which is detected as the spam properly; in other words, the record of test data with the original class 1 is put in the Class 1 by the classifier.

FP: The valid email, which is predicted as the spam by mistake; in other words, the record of test data with the original class 0 is put in the Class 1 by the classifier.

TN: The valid email, which is predicted as the valid email properly; in other words, the record of test data with the original class 0 is put in the Class 0 by the classifier.

FN: The spam, which is predicted as the valid email by mistake; in other words, the record of test data with the original class 1 is put in the Class 0 by the classifier.

$$Precision - Class\ 1 = \frac{TP}{(TP+FP)} \quad (1)$$

$$Recall - Class\ 1 = \frac{TP}{(TP+FN)} \quad (2)$$

$$F - Measure = \frac{Precision * Recall}{Precision + Recall} \quad (3)$$

$$Precision - Class\ 0 = \frac{TN}{(TN+FN)} \quad (4)$$

$$Recall - Class\ 0 = \frac{TN}{(TN+FP)} \quad (5)$$

Table 1: Results obtained from the class processing of Spam by involving all features

|  | TP | FP | Precision | Recall | F-Measure |
|---|---|---|---|---|---|
| SVM | 0.757 | 0.024 | 0.959 | 0.757 | 0.846 |
| Naive Bayesian | 0.963 | 0.231 | 0.757 | 0.963 | 0.848 |
| ADTree | 0.958 | 0.075 | 0.957 | 0.925 | 0.940 |
| LADTree | 0.955 | 0.112 | 0.952 | 0.955 | 0.924 |
| J48 | 0.951 | 0.112 | 0.948 | 0.951 | 0.922 |
| **Voting** | 0.937 | 0.07 | **0.937** | 0.93 | 0.933 |

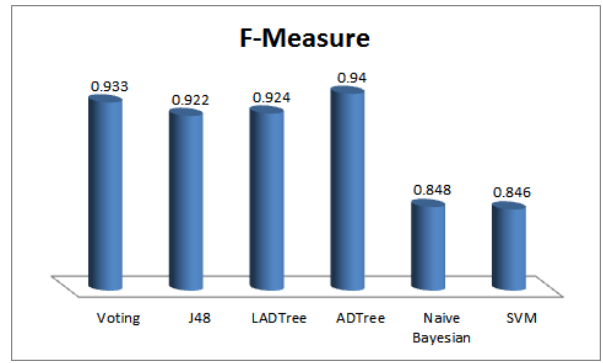

Fig. 6 Comparison of F-Measure, The results obtained from processing the spam class with involving all features

Table 2: Results obtained from processing the class of valid email

|  | TN | TN | Precision | Recall | F-Measure |
|---|---|---|---|---|---|
| SVM | 0.976 | 0.243 | 0.843 | 0.976 | 0.904 |
| Naive Bayesian | 0.769 | 0.037 | 0.965 | 0.769 | 0.856 |
| ADTree | 0.925 | 0.042 | 0.927 | 0.958 | 0.942 |
| LADTree | 0.888 | 0.045 | 0.895 | 0.888 | 0.918 |
| J48 | 0.888 | 0.049 | 0.895 | 0.888 | 0.917 |
| Voting | 0.941 | 0.065 | 0.935 | 0.941 | 0.937 |

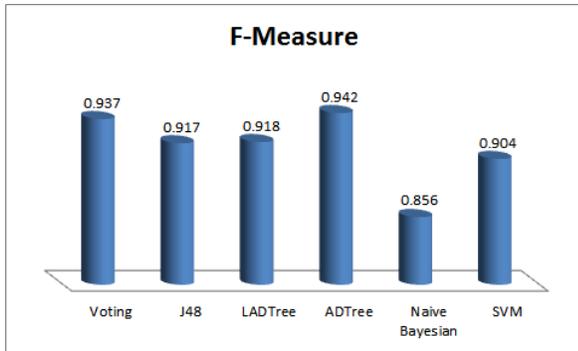

Fig. 7 Comparison of F-Measure obtained from processing the class of valid email with involving all features

As we can see, our proposed method has been able to be more successful than the previous methods in both classes of spam and valid email.

## 6- Conclusion and Future Work

In this paper, we have Used Weka software to produced the decision trees AD, LAD and J48 and and have converted them into the format of ontology (RDF) and ensured their accuracy, and then the test emails are given to them and we have voted for the predicted results and finally we have compared the obtained results with the results of two methods SVM and Naive Bayesian which are the most common Email classification methods; thus we have found that the results obtained from voting the decision trees between two errors of considering the spam instead of valid email (FN) and a valid email instead of spam (FP) establish a reasonable balance.

Pruning and narrowing the constructed tree as a basis of the proposed ontology can be done as the future work. Furthermore, the classifiers such as Conjunctive-Rule and other non-tree classifiers can be used as well as the tree classifiers in voting step in order to achieve the higher accuracy.